\documentclass[]{aa}
\usepackage{graphicx}
\usepackage{txfonts}
\begin{document}

\title{The magnetic field of the pre-main sequence Herbig Ae star
HD 190073
\thanks{Based on observations obtained at the Canada-France-Hawaii
Telescope (CFHT) which is operated by the National Research Council 
of Canada, the Institut National des Sciences de l'Univers of the
Centre National de la Recherche Scientifique of France, and the 
University of Hawaii}
}

\author{
C. Catala\inst{1} \and
E. Alecian\inst{1} \and
J.-F. Donati\inst{2} \and
G.A. Wade\inst{3} \and
J.D. Landstreet\inst{4} \and
T. B\"ohm\inst{2} \and
J.-C. Bouret\inst{5} \and
S. Bagnulo\inst{6} \and
C. Folsom\inst{3,7} \and
J. Silvester\inst{3,7}
}

\institute{
Observatoire de Paris, LESIA, CNRS UMR 8109, 5, place Jules Janssen, F-92195
Meudon Principal CEDEX, France
  \and
Laboratoire d'Astrophysique, Observatoire Midi-Pyr\'en\'ees,
14 avenue Edouard Belin, F-31400 Toulouse, France
  \and
Dept. of Physics, Royal Military College of Canada, PO Box 17000, Stn Forces,
Kingston, Canada K7K 4B4
  \and
Dept. of Physics \& Astronomy, University of Western Ontario, London, Canada
N6A 3K7
  \and
Laboratoire d'Astrophysique de Marseille, Traverse du Siphon,
BP8-13376 Marseille Cedex 12, France
  \and
European Southern Observatory, Casilla 19001, Santiago 19, Chile
  \and
Department of Physics, Engineering Physics, and Astronomy, 
Queen's University, Kingston, ON K7L 3N6, Canada
}

\date{\today}

\offprints{C. Catala (claude.catala@obspm.fr)}

\date{Received , accepted }

\abstract
{The general context of this paper is the study of magnetic fields in the 
pre-main sequence intermediate mass Herbig Ae/Be stars. Magnetic fields are 
likely to play an important role in pre-main sequence evolution at 
these masses, in particular in controlling the gains and
losses of stellar angular momentum.}
{The particular aim of this paper is to announce the detection of a
structured magnetic field in the Herbig Ae star HD 190073, and to discuss
various scenarii for the geometry of the star, its environment and its
magnetic field.}
{We have used the ESPaDOnS spectropolarimeter at CFHT in 2005 and 2006 to 
obtain high-resolution and signal-to-noise circular polarization spectra which 
demonstrate unambiguously the presence of a magnetic field in the photosphere 
of this star.}
{Nine circular polarization spectra were obtained, each one showing a clear 
Zeeman signature. This signature is suggestive of a 
magnetic field structured on large scales. The signature, which corresponds 
to a longitudinal magnetic field of $74\pm 10$ G, does not vary detectably on 
a one-year timeframe, indicating either an azimuthally symmetric field, 
a zero inclination angle between the rotation axis and the line of 
sight, or a very long rotation period. 
The optical spectrum of HD 190073 exhibits a large number of emission
lines. We discuss the formation of these emission lines in the framework of
a model involving a turbulent heated region  at the base of the
stellar wind, possibly powered by magnetic accretion.} 
{This magnetic detection brings an important element for our understanding
of stellar magnetism at intermediate masses.}


\titlerunning{The magnetic field of HD 190073}

\maketitle

\section{Introduction}
Magnetic fields can have a strong impact on pre-main sequence (PMS) evolution. 
In particular, the role of magnetic fields in the evolution of angular momentum
during the PMS phase can be crucial, both by amplifying the angular momentum 
losses through stellar winds, and by coupling the stars with the massive 
accretion disks present around a large fraction of them. Magnetic fields 
are probably also important for evacuating angular momentum from the disks
themselves in very early phases of star formation, impacting the initial 
conditions for pre-main sequence evolution.

In the case of the low mass PMS stars, the T Tauri stars, stellar magnetic 
fields are believed to channel the accretion flow toward the star's surface
along magnetic lines, and to control the accretion disk structure in this
magnetospheric accretion region (K\"onigl 1991, Paatz \& Camenzind 1996). 
The topology of this channelled accretion depends strongly on the magnetic
field structure and the tilt angle between rotation and magnetic
axis (Romanova et al. 2003).
While evidence has been accumulated
recently that a large fraction of the intermediate mass PMS stars, the 
Herbig Ae/Be stars, also host massive accretion disks (Grady 2005), 
the magnetospheric accretion scenario has not been investigated in detail 
in their case.

The search for magnetic fields in the Herbig Ae/Be stars and, if they exist, 
the detailed analysis of their strength and topology, are necessary steps 
in the study of PMS evolution and the interplay between the stars, their 
accretion phenomena and their winds.

Another fundamental reason for investigating magnetic fields in Herbig stars 
is related to our understanding of the strong globally-ordered magnetic fields
observed in the chemically peculiar A and B stars (Ap/Bp stars). The most
common interpretation, known as the primordial fossil field hypothesis, 
involves a magnification of interstellar magnetic field swept up during 
the process of star formation (Mestel 2001, Moss 2001). In this hypothesis, 
we expect to observe a fraction of the Herbig Ae/Be stars which also host 
globally-ordered magnetic fields, representing the progenitors of the 
magnetic Ap/Bp stars.
 
Most Herbig Ae/Be stars show conspicuous signs of winds and activity. These
active phenomena have often been assumed to be magnetic 
(Praderie et al. 1986, Catala et al. 1999), but no direct nor 
undisputable proof of this magnetic origin has been presented so far.

Spectropolarimetric observations have recently provided some new insight
into the problem of magnetic fields of Herbig Ae/Be stars. After the 
pioneering measurement of an effective longitudinal field of about
50~G in HD 104237 by Donati et al. (1997), using the UCLES spectrograph on the
AAT, equipped with the visitor SemelPol polarimeter, Hubrig et al. (2004) 
announced the discovery of a magnetic field in the Herbig Ae star
HD 139614, as well as 2 other marginal detections, using FORS1 in polarimetric
mode on the VLT. This result was not confirmed by more recent, higher
S/N ratio observations with the ESPaDOnS spectropolarimeter at CFHT, 
which on the other hand provided clear magnetic detections for the 
Herbig Ae stars V380 Ori and HD 72106A (Wade et al. 2005). These authors
also report the detection of a field in HD 101412, using FORS1 on the VLT.

Our knowledge of magnetic fields in Herbig stars is evolving rapidly. 
This paper constitutes another step in that direction, by presenting 
results of spectropolarimetric observations of one more Herbig star,
HD 190073. This early-type star (A2IIIe-B9IVep+sh, Pogodin
et al. 2005) with many emission lines in its visible spectrum has been
recently shown to be a young Herbig Ae/Be star (Cidale et al. 2000, de Winter 
et al. 2001). Its spectral energy distribution in the infrared, as well as
its 9.7 $\mu$m silicate feature, are very similar to those of well-known 
Herbig stars (Malfait et al. 1998, Chen et al. 2000). Its remarkably low
projected rotational velocity ($v \sin i = 9$ km\thinspace s$^{-1}$, 
Acke \& Waelkens 2004) is rather unusual among Herbig Ae/Be stars, and can be
indicative of either a very slow rotation, or a very small inclination of the
rotation axis with respect to the line of sight. Recent interferometric 
observations in the infrared are best interpreted in terms of a circumstellar
disk seen nearly face-on, although higher inclination angles cannot be ruled
out by these observations, nor the interpretation of the interferometric 
visibilities as due to binarity (Eisner et al. 2004).

In Sect. 2, we present our spectropolarimetric observations and data 
reduction procedures. 
The results of our investigations are presented in Sect. 3. A conclusion 
is given in Sect. 4. 

\section{Observations and data reduction}

We used the ESPaDOnS spectropolarimeter, recently installed on the 
3.6m Canada-France-Hawaii Telescope (Donati et al. 2006, in preparation), during
3 observing runs in 2005 partly devoted to Herbig Ae/Be stars. Table 1 presents
the log of the observations.

\begin{table}
\caption{Journal of ESPaDONS observations of HD 190073. 
The 5th column gives the S/N ratio 
at 600 nm per spectral bin of 0.035 nm, while the 6th column lists the 
S/N ratio in the deconvolved LSD Stokes $V$ profile per velocity bin of 
1.8 km\thinspace s$^{-1}$. The last column lists the resulting effective
longitudinal magnetic field, calculated from the LSD profile using the 
complete line mask, as discussed in Sect. 2 and 3.3.}
\scriptsize
\begin{tabular}{lllllll}
\hline
  Date       &   JD         &  UT   & $t_{\rm exp}$ & S/N & S/N  & $B_{eff}$ \\
dd/mm/yy   & (2,453,000+) &       &    (s)    &     & (LSD) &   (G)    \\
\hline
22/05/05   & 512.989      & 11:40 & 3290      & 370 & 3600 &$+69 \pm 15$ \\
23/05/05   & 513.969      & 11:11 & 3600      & 250 & 2500 &$+61 \pm 35$ \\
24/05/05   & 514.960      & 10:59 & 3600      & 400 & 3950 &$+68 \pm 17$ \\
24/05/05   & 515.065      & 13:59 & 2400      & 300 & 3100 &$+66 \pm 22$ \\
25/05/05   & 515.962      & 11:01 & 3600      & 370 & 3700 &$+64 \pm 16$ \\
25/05/05   & 516.088      & 14:02 & 2400      & 290 & 2950 &$+98 \pm 30$ \\
\hline
19/07/05   & 570.916      & 09:51 & 1800       & 470 & 5300 &$+70 \pm 11$ \\
\hline
25/08/05   & 607.789      & 06:49 & 2000       & 500 & 5500 &$+74 \pm 10$ \\
\hline
08/06/06   & 895.921      & 10:01 & 2400       & 560 & 5900 &$+73 \pm 10$ \\
\hline 
\end{tabular}
\end{table}

The data were obtained in the polarimetric configuration of ESPaDOnS,
yielding a spectral resolution of 65,000. All spectra were recorded as 
sequences of 4 individual subexposures taken in different configurations of 
the polarimeter, in order to yield a full circular polarization analysis, 
as described in Donati et al. (1997) and Donati et al. (2006, in preparation). 
No linear polarization analysis was performed. The data were reduced with 
the automatic reduction package "Libre-ESpRIT" installed at CFHT (Donati et 
al.  1997, Donati et al. 2006 in preparation). Stokes $I$ and Stokes $V$
spectra are obtained by proper combinations of the 4 subexposures, while
check spectra, labelled as $N$ spectra, are calculated by combining
the subexposures in such a way to yield a null signal, that can be
used to verify the reality of the signal measured in Stokes $V$.

The data of May 2005 are affected by a 1.3 mag loss compared to the other
data presented here. This problem, which was due to damage to the external
jacket of optical fibres, was fixed prior to the July run.

We subsequently applied the Least-Square Deconvolution (LSD) method described
in Donati et al. (1997) to construct average photospheric profiles both 
of the $I$ and $V$ Stokes parameters. The LSD technique builds the average 
photospheric line profile by deconvolving the observed spectrum (both in $I$
and $V$ Stokes parameters, as well as for the null $N$ spectrum) 
from a line mask including all lines present in
a synthetic spectrum of the star.  The line mask was computed using a Kurucz 
Atlas 9 model with effective temperature, surface gravity and 
metallicity adequate for HD 190073, compiled in Table 2. 
In the deconvolution procedure, each line is weighted by the product of
its S/N ratio in 
the observed spectrum, its depth in the unbroadened Kurucz model, and its 
magnetic Land\'e factor. Hydrogen Balmer lines, strong He I lines, strong 
resonance lines, the Ca II IR triplet lines, as well as lines for which 
the magnetic Land\'e factor cannot be computed, were excluded from the 
mask, which otherwise contains all other lines in the model whose depth relative
to the continuum is larger than 0.1. In addition to this main line mask, 
we also constructed a sub-mask 
containing only lines whose relative depth in the Kurucz model is comprised
between 0.1 and 0.4.
We shall see later (Sect. 3.2) that these shallow lines have no 
or only very weak emission components, and this mask will therefore be 
used to study the purely photospheric contribution to the spectrum.
The complete mask contains 1,400 lines in the ESPaDOnS spectral domain, while
the shallow-line mask has 1,000 lines.

The LSD average line profiles were computed on a velocity grid with 
a 1.8 km\thinspace s$^{-1}$ sampling. The resulting relative noise in the
LSD $V$ Stokes profiles is given in the 6th column of Table 1.

Finally, both non-magnetic and magnetic standard stars were observed with
ESPaDOnS, allowing us to verify the 
nominal behaviour of the instrument (see Donati et al. 2006, in preparation, 
for details).

\section{Results}

\subsection{Fundamental parameters of HD 190073}

The effective temperature and surface gravity of HD 190073 
were taken from Acke \& Waelkens (2004), as T$_{\mathrm{eff}}$~=~9,250~K and
log~$g$~=~3.5 (cf Table 2). The determination of its bolometric 
luminosity is made difficult by the fact that the Hipparcos parallax has
a large error bar (Van den Ancker et al. 1998). Acke et al. (2005)
adopt Log$(L/L_{\odot}) = 1.92$, which is compatible with the lower limit
of 1.80 determined by Van den Ancker et al. (1998). In this paper, we shall 
adopt the same value as Acke et al. (2005), and a $\pm 0.12$ error bar to 
account for these difficulties.

\begin{figure}
\centering
\includegraphics[width=9cm]{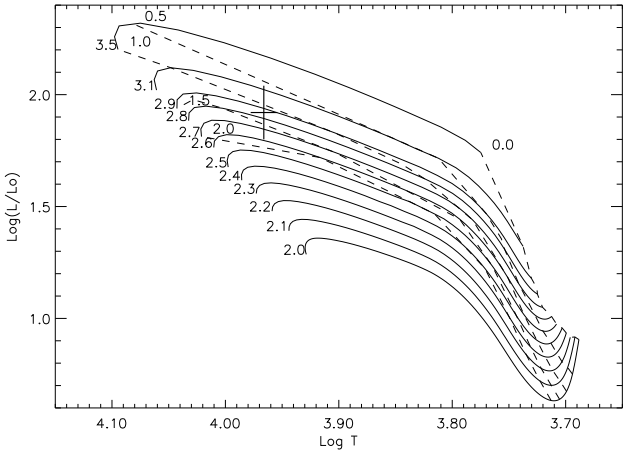}
\caption[]{The location of HD 190073 in the HR diagramme and the corresponding
error bar are shown as a cross. Evolutionary tracks computed by CESAM are 
depicted in full lines and labelled by their masses. 
Isochrones are shown as dashed lines and labelled in Myrs. Pre-main sequence
evolutionary tracks start on the birthline and ages are counted using the
birthline as time origin.} 
\label{hrd}
\end{figure}

We compared the location of HD 190073
in the HR diagramme to theoretical evolutionary tracks computed with the CESAM 
code (Morel 1997). This location is shown in Fig.~\ref{hrd}, where
evolutionary tracks for masses between 2 and 3.5 $M_{\odot}$, and
isochrones from 0 to 2 Myrs are shown. We assumed that PMS evolution
started on the birthline in the HR diagramme, as defined by Palla \& Stahler
(1990). Stellar ages are counted on each track using the birthline as time
origin. Comparison of the location of HD 190073 in the HR diagramme with 
evolutionary tracks and isochrones yields estimates of its mass 
(2.85 $M_{\odot}$) and age (1.2 Myrs), and
the corresponding CESAM models can be used to estimate its radius, 
$R = 3.6 R_{\odot}$. The mass and radius of HD 190073 derived in this way
yield a surface gravity Log $g = 3.7$, to be compared to the value
Log $g = 3.5$ derived by Acke \& Waelkens (2004) from spectroscopy.

\begin{figure}
\centering
\includegraphics[width=9cm]{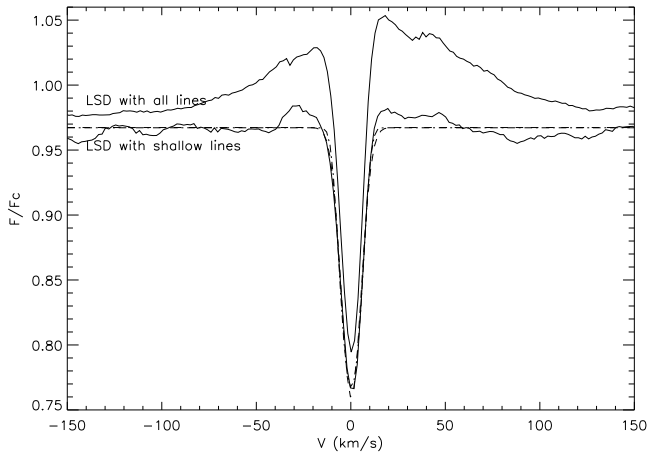}
\caption[]{The LSD average line profile of HD 190073 in August 2005. Average 
line profiles
computed with both types of line masks described in the text are shown. 
The dashed line shows the gaussian + rotation fit to the profile, while the
dashed-dotted line corresponds to the purely gaussian fit. Both models yield 
almost identical profiles.}
\label{lsd_profile}
\end{figure}

As described in Sect. 3.2, the spectrum of HD 190073 is 
characterized by the presence of numerous emission lines, on which photospheric
absorption lines are superimposed. When performing the LSD deconvolution to 
determine the average line profile, in addition to the normal line mask, we 
also used the sub-mask described in the
previous section, containing only shallow lines with no or moderate 
emission, in order to study the photospheric line profiles without being 
perturbed by emission. Figure~\ref{lsd_profile} shows the average line profiles
computed using both masks, using the spectrum recorded in August 2005.
We fitted this profile both with a simple gaussian
and with the convolution of a gaussian and a rotation profile, calculated 
following Gray (1992). In the second case, the gaussian involved in the 
convolution included contributions from instrument broadening, thermal 
broadening, and turbulence. We used an instrument broadening 
corresponding to a resolving power $R = 65,000$, adequate for ESPaDOnS, 
and a turbulent velocity of 2 km\thinspace s$^{-1}$. 

Both types of fit are shown in Fig.~\ref{lsd_profile}. The gaussian best 
fit yields a total line FWHM of 
12.4 km\thinspace s$^{-1}$, corresponding to a 
total turbulent velocity v$_{turb}$ of 
6.8 km\thinspace s$^{-1}$ 
when instrumental broadening is taken into account (v$_{turb}$
is defined throughout the paper as yielding a gaussian broadening proportional 
to exp$[$-v$^2$/v$_{turb}^2]$). 
The gaussian+rotation fit implies a projected rotation
velocity of 8.6 km\thinspace s$^{-1}$. We argue that both solutions are 
equally compatible with the data.  

\begin{figure}
\centering
\includegraphics[angle=-90,width=9cm]{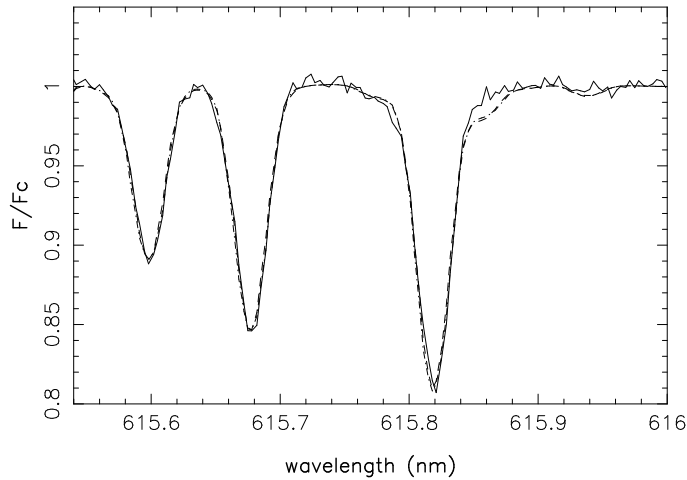}
\caption[]{The spectral region of the photospheric O~I lines in August 2005. 
The solid line represents the observed profile, while the dashed line 
corresponds to the model with no rotation and a macroturbulent velocity
of 6 km\thinspace s$^{-1}$, and the dashed-dotted line to a model with
no macroturbulence and a projected rotation velocity of 
8.6 km\thinspace s$^{-1}$. Both models yield almost identical profiles.}
\label{oi_lines}
\end{figure}

We note however that even the LSD profile
constructed with the shallow-line mask includes some lines with a significant
emission component, as well as lines where the photospheric absorption
is filled-in by some emission, which can perturb the analysis of the line 
broadening. 
We therefore examined in detail a few lines which appear purely photospheric
with no emission at all. The best examples of such lines are the O I lines 
near 616 nm. In Fig.~\ref{oi_lines}, we compare the profile of these lines
with Kurucz Atlas 9 synthetic spectra, computed assuming the effective 
temperature and surface gravity listed in Table 2, and using detailed
chemical abundances as determined by Acke \& Waelkens (2004). We computed 
two different series of
Kurucz spectra: the first one includes a microturbulent velocity of 
2 km\thinspace s$^{-1}$, an instrumental broadening corresponding to 
the spectral resolution of ESPaDOnS, and a rotational broadening
left as a fitting variable; the second series of Kurucz spectra includes no
rotation, a microturbulent velocity of 2 km\thinspace s$^{-1}$, and an 
isotropic macroturbulent velocity left as free variable. We find that the 
O~I lines can equivalently be fitted either by a rotating model with 
$v \sin i$ = 8.6 km\thinspace s$^{-1}$ and no macroturbulence, or by a model
including a macroturbulent velocity of 6.0 km\thinspace s$^{-1}$ and no
rotation. Appropriate combinations of rotation and macroturbulence with 
intermediate values would of course fit the data as well.

\begin{figure*}[!h]
\centering
\includegraphics[width=14cm,height=7cm]{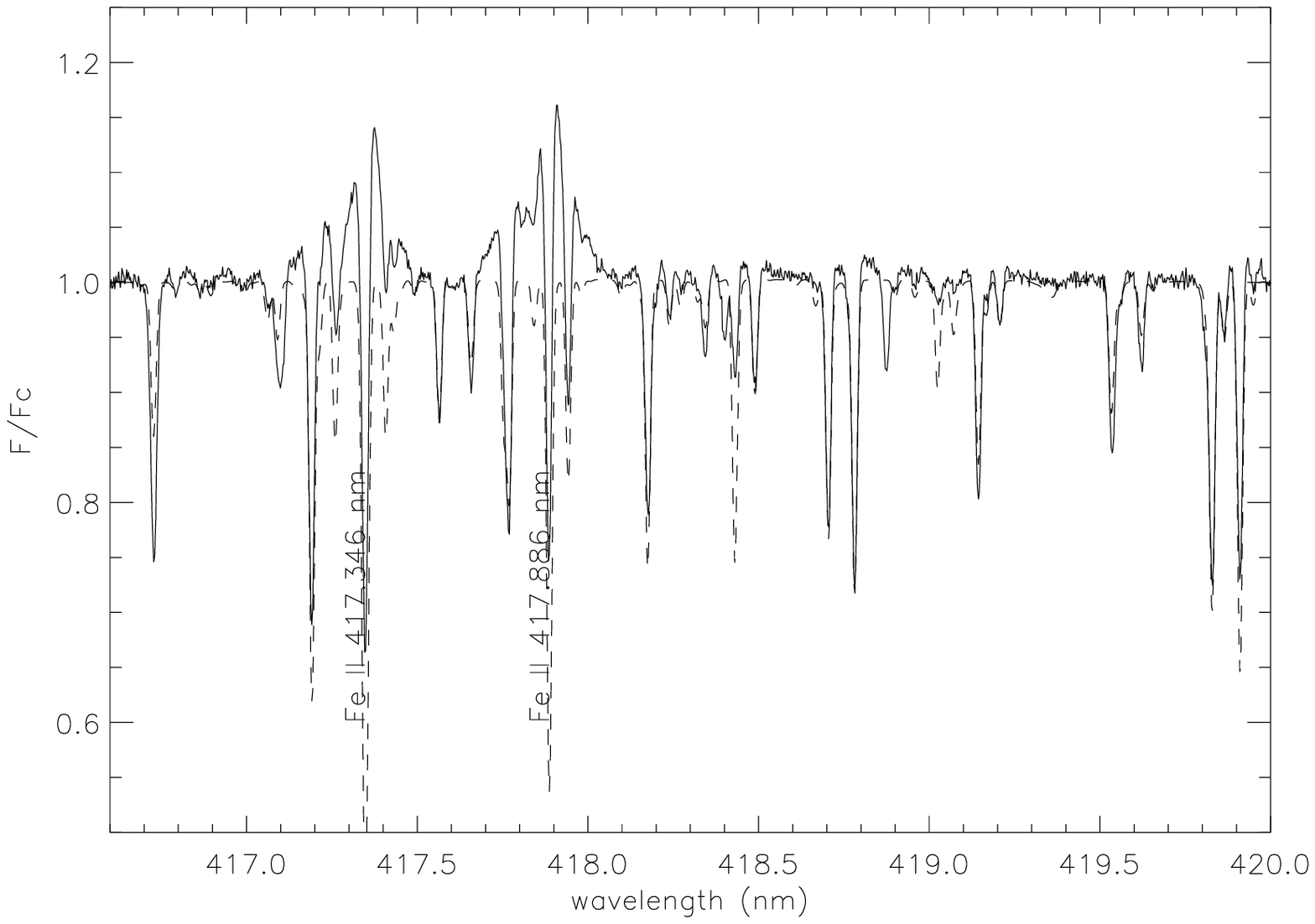}
\includegraphics[width=14cm,height=7cm]{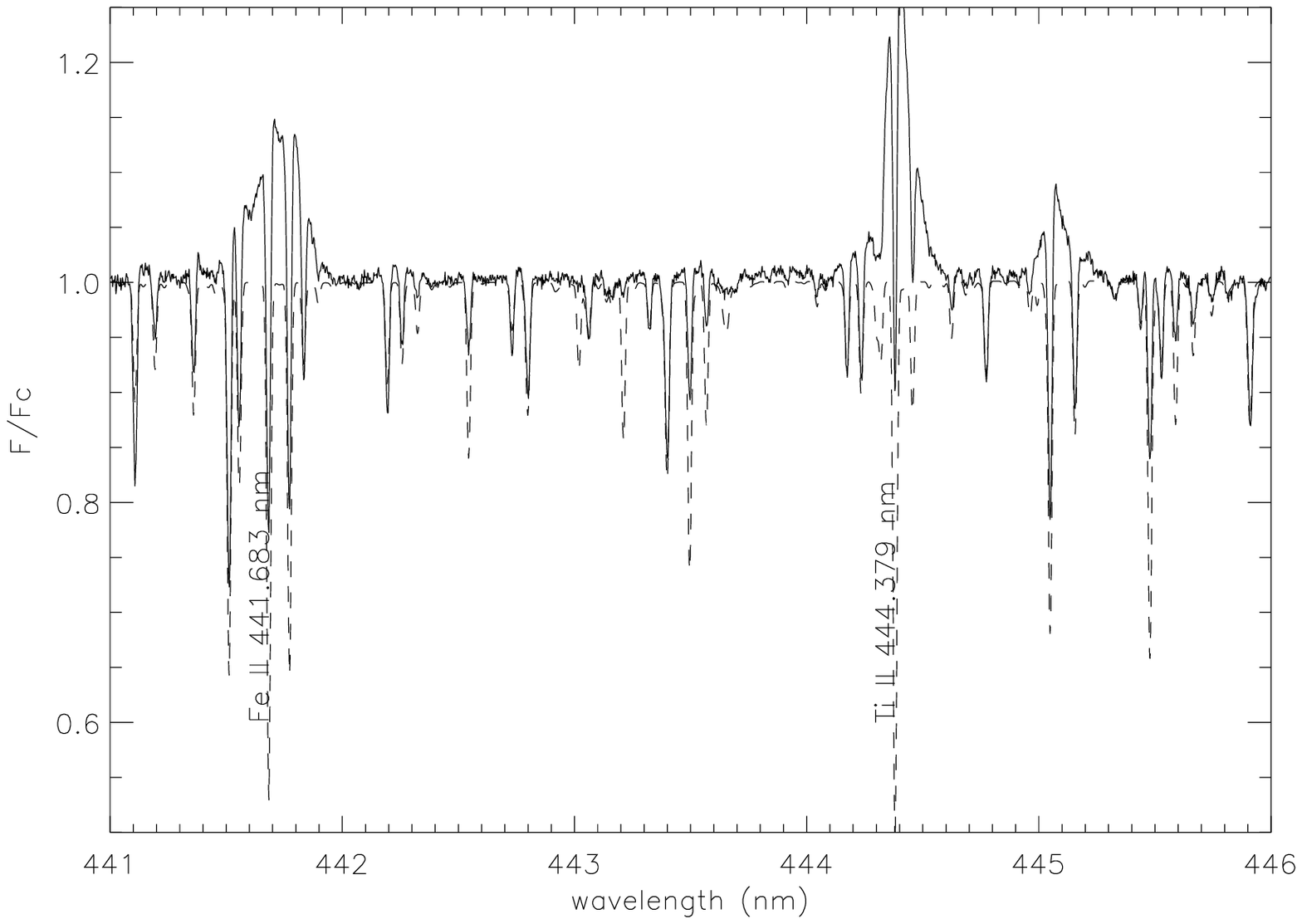}
\includegraphics[width=14cm,height=7cm]{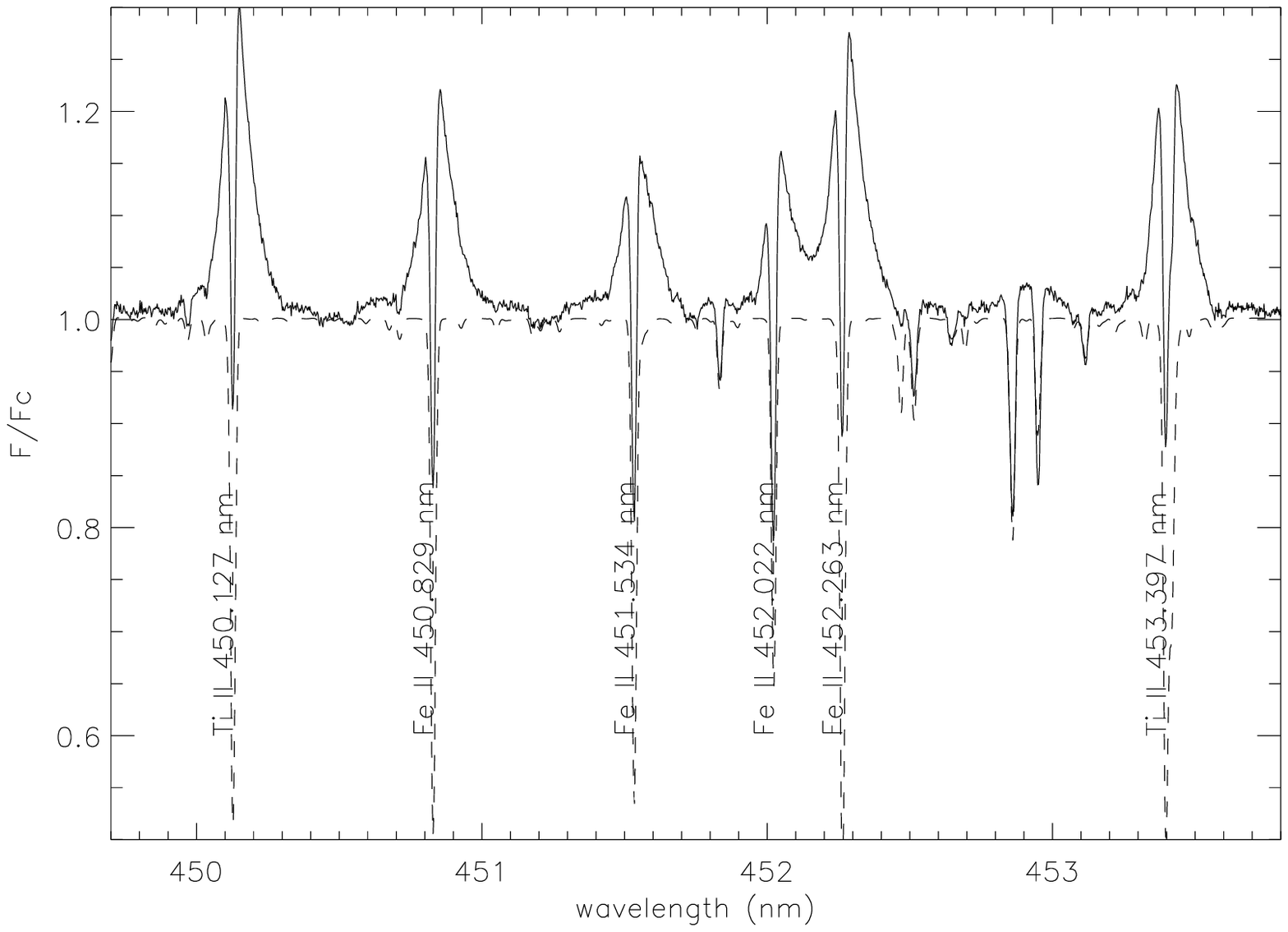}
\caption[]{Portions of the optical spectrum of HD 190073 in August 2005
(full line),
compared to a synthetic spectrum computed from the Kurucz Atlas 9 model with
T$_{\mathrm{eff}}$=9,250 K, log $g$=3.5, $[Fe/H]=0$,
$v \sin i$ = 0 km\thinspace s$^{-1}$, $V_r = 0.21$ km\thinspace s$^{-1}$,
$V_{micro}$ = 2.0 km\thinspace s$^{-1}$ and
$V_{macro}$ = 6.0 km\thinspace s$^{-1}$
(dashed line). Note that faint lines are reasonably well reproduced by the 
synthetic model, while stronger lines show emission components.}
\label{hd190073_I}
\end{figure*}

We conclude from the above analysis that both rotating and macroturbulent
models can equally reproduce the O~I line profiles. This conclusion and 
the derived values of the projected rotation and macroturbulent velocities 
are in fair agreement with
those derived from the analysis of the LSD profile. The examination of other 
lines in the spectrum led us to similar conclusions. Therefore, the projected 
rotation velocity can be anywhere between 0 and 8.6 km\thinspace s$^{-1}$, 
while the macroturbulent velocity can range from 0 to 6.0 km\thinspace s$^{-1}$.
Although such high macroturbulent velocities may be surprising in an 
early A-type star like HD 190073, we note that significant photospheric 
turbulent velocities have been reported for main sequence A-type stars up to 
spectral type A0 (Landstreet 1998). LTE atmospheres are unstable to 
convection up to 
about T$_{\mathrm{eff}}$=10,000 K, due to the high opacity 
of excited and partly ionized hydrogen. Although this convection, which carries
negligible heat, has no effect on the atmospheric structure, it still occurs
and may result in photospheric turbulent velocity fields. In addition, 
accretion and mass loss processes which are known to occur in very young 
stars like HD 190073 may also lead to additional photospheric turbulence. 
Hence the possibility of having photospheric turbulent velocities as high 
as 6 km\thinspace s$^{-1}$ 
in this star should be considered seriously.

We also attempted to use the Fourier transform of line profiles 
to disentangle turbulent velocity from rotation, following Reiners \& Schmidt 
(2003). We calculated the Fourier transform of the LSD  average profile, as
well as that of individual unblended photospheric lines, such as the O~I lines
discussed earlier. Unfortunately, we find that $v \sin i$ values lower than 
9 km\thinspace s$^{-1}$ cannot be distinguished by this method, neither
with the LSD profile nor with individual profiles, even 
with a high signal to noise ratio as in our data. 

All derived fundamental parameters are summarized in Table 2.

\begin{table*}
\caption{Fundamental parameters of HD 190073. See text for the range of 
$v \sin i$ and v$_{macro}$ values.}
\begin{tabular}{lllllllll}
\hline
T$_{\mathrm{eff}}$ & log $g$  & Log$(L/L_{\odot})$  &    
$M/M_{\odot}$  & $R/R_{\odot}$ & age    & $v \sin i$ & v$_{\rm macro}$ & v$_{\rm rad}$ \\
(K)  & (cm\thinspace s$^{-2}$)&                    &  
               &               & (Myr)  & (km\thinspace s$^{-1}$) & (km\thinspace s$^{-1}$) & (km\thinspace s$^{-1}$)  \\
\hline
$9,250 \pm 250$  & $3.5 \pm 0.5$   & $1.92 \pm 0.12$      & 
  $2.85 \pm 0.25$ &  $3.6 \pm 0.5$  & $1.2 \pm 0.6$       &   0 -- 8.6 & 0 -- 6.0 & 0.21 $\pm 0.1$ \\
\hline
\end{tabular}
\end{table*}

\subsection{Unpolarized spectrum}

We observed repeatedly the Herbig Ae star HD 190073 with ESPaDOnS between
May 2005 and June 2006. The star was observed up to twice a night for 4 
consecutive nights in May, then additional isolated spectra were obtained in
July and August 2005 and in June 2006 (see Table 1 for the log of 
observations). This observing strategy allows us to assess variability on 
a daily, monthly and yearly basis.
The spectroscopic behaviour of HD 190073 was thoroughly studied by Pogodin 
et al. (2005), and we basically confirm their findings with our new 
observations. 

The optical spectrum of HD 190073 is globally that of a star with
T$_{\mathrm{eff}}$=9,250 K and log $g$=3.5, including a large number
of emission lines. As a matter of fact, we find that many lines of ions such
as Ca~I, Ca~II, Fe~I, Fe~II, Ti~II, Si~II, Sc~II, Cr~II, present in the 
Atlas 9 Kurucz synthetic spectrum adequate for HD 190073, having an intrinsic 
depth larger than 0.4 (i.e. the depth of the line in the synthetic spectrum
in the absence of any broadening) exhibit an emission in HD 190073. 

All of these emission lines are overlapped by a narrow absorption, with a
width similar to that of the photospheric lines with no emission. 
The absorption lines and the absorption components overlapping the emission
lines are in good agreement with the Kurucz synthetic spectrum.
We therefore consider that these absorptions are of photospheric origin rather 
than due to a shell as discussed in Pogodin et al. (2005). 
The detailed analysis of Acke \& Waelkens (2004) indicates chemical abundances
pretty close solar, and our own synthesis using these
abundances are in very good agreement with our observations, as shown for
example by the fit of the O~I lines in Fig.~\ref{oi_lines}. This star
therefore does not show the usual strong chemical peculiarities of Ap stars.

The emission components themselves have the same 
width all across the spectrum, equal to 65 $\pm 7$ km\thinspace s$^{-1}$ 
(FWHM). The intensity of these emission lines exhibit some moderate 
variability, of the order of a couple of percent in the one-year
timeframe of our observations, while keeping the same shape and width. 
We also note that the emission lines have a centroid significantly 
redward of the absorption components that are superimposed on them, 
with a velocity variable between 4 and 19 km\thinspace s$^{-1}$, in
an irregular manner from one epoch to another.
Figure~\ref{hd190073_I} shows an extract of the optical
spectrum of HD 190073, compared to a Kurucz synthetic spectrum.

\begin{figure}
\centering
\includegraphics[width=9cm]{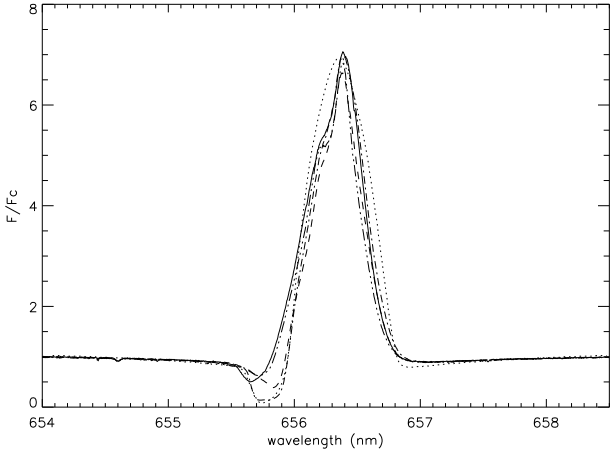}
\caption[]{The H$\alpha$ P Cygni line profile of HD 190073 
May 2005 (full line), 
July 2005 (dashed line), August 2005 (dashed-dotted line) and June 2006 
(dashed-dot-dot-dotted line). Note the
strong variation of the absorption component on a monthly basis, while the
emission is only moderately variable. The computed profile fitted to the
August 2005 data is shown as a dotted line.}
\label{halpha}
\end{figure}

We also confirm the presence of strong P Cygni profiles at H$\alpha$ and other
hydrogen Balmer lines, with a strong monthly variability: between May and 
August 2005, although the red emission of these profiles remains moderately 
variable, we see a strong increase in the blueward absorption component. 
By June 2006, the absorption component of H$\alpha$ has come back to its
shape of May 2005. In contrast to this strong variability, the H$\alpha$
emission component shows only moderate variations.
This variability is shown in Fig.~\ref{halpha}.

The most natural interpretation the H$\alpha$ P Cygni profile of
HD 190073 involves a stellar wind.
We have tentatively modelled this profile with a spherically
symmetric wind model, similar to that of Bouret \& Catala (1998). This model
includes a heated region at the base of the wind, surrounded by
a cooler wind extending up to several tens of stellar radii. The coupled
radiative transfer and statistical equilibrium equations for the hydrogen
atom were solved in the comoving frame of the flow, using the 
Equivalent Two-Level Approach (ETLA) introduced by Mihalas \& Kunasz (1978) 
and used in Bouret \& Catala (1998). We find that the H$\alpha$ P Cygni
profile observed in August 2005 can be satisfactorily modelled with a
wind having a mass loss rate $\dot{M}$ of 
$1.4 \pm 0.3 \times 10^{-8}$ M$_{\odot}$yr$^{-1}$
and a terminal velocity of $290 \pm 10$ km\thinspace s$^{-1}$. 
The temperature of the
model in the heated region at the base of the wind reaches 
18,000 K, although we cannot constrain it tightly with the H$\alpha$ 
profile. The H$\alpha$ line profile 
computed with this model in shown in Fig.~\ref{halpha}. It fits reasonably
well the August 2005 observed profile. We have not attempted to model 
the strong variability of the observed absorption component, which is 
presumably due to the presence of structures in the wind appearing on the
line of sight. On the other hand, the global properties of the wind, e.g. 
the mass loss rate, remain constant, producing a more or less 
stable emission component for this line (see Bouret \& Catala 1998, for 
a more detailed discussion). Note finally that the estimate of the 
mass loss rate given above assumes that all of the H$\alpha$ emission 
originates from the wind. While the existence of a stellar wind is clearly
demonstrated by the P Cygni profile at H$\alpha$, other processes, such as 
emission from hot regions above the photosphere can contribute to this 
emission (see below), so that our estimate of $\dot{M}$ should rather be
considered as an upper limit.

Quite in contrast with the H$\alpha$ line, the Ca II K \& H lines in our 
spectra do not show conspicuous varibility in our one year timeframe, and 
are even very similar to the profiles presented by Pogodin et al. (2005), 
observed from 1994 to 2000.

\subsection{Magnetic field}

\begin{figure}
\centering
\includegraphics[height=17cm,totalheight=18cm]{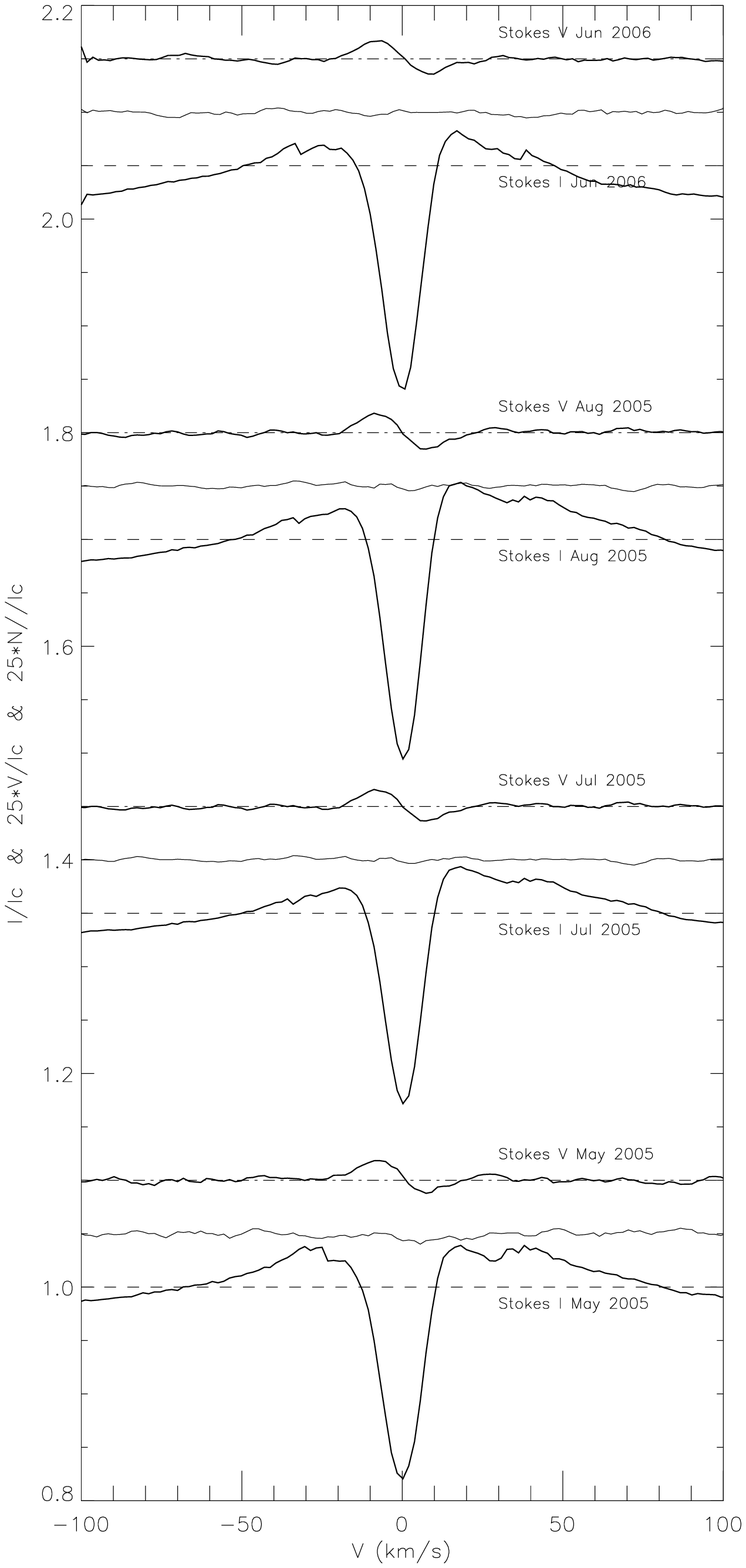}
\caption[]{LSD unpolarised and circularly polarised profiles of HD 190073,
at various epochs in 2005 and 2006. 
All LSD profiles are calculated with the mask including all photospheric lines. Spectra are shifted for display purposes.
The curves in light line appearing between the Stokes $I$ and $V$ profiles
are the "null" spectra calculated as described in Sect. 2.
The Stokes $V$ profiles and the null $N$ profiles 
are expanded by a factor 25. The profile labelled "May 2005" is that of May 24,
10:59 UT.
All spectra are in heliocentric restframe.}
\label{hd190073_V}
\end{figure}

We detected a conspicuous Stokes $V$ signature of HD 190073, at all 
epochs in 2005 and 2006. This signature in $V/I$ has an amplitude of about 
$1 \times 10^{-3}$, and is presented in Fig.~\ref{hd190073_V}. The shape of 
this Stokes $V$ signature is very simple, and indicates a globally structured
magnetic field.  The observed signature remains remarkably constant
in all our observations from May 2005 to June 2006, and can be translated to 
an effective longitudinal magnetic field $B_{eff}$ of $+74 \pm 10$ G, 
as measured on the August 2005 spectrum, using the approach described in 
Eq. 5 of Donati et al. (1997). 
The $\pm 10$ G error bar is calculated from the noise
level of the August 2005 LSD Stokes $V$ profile, and the effective longitudinal
magnetic fields measured from the other spectra of HD 190073 remain in agreement
with this error bar. All $B_{eff}$ measurements are compiled in Table 1. 

We also calculated the standard deviation of the
Stokes $V$ profile in time, for each velocity bin  across the profile.
We found it to be independent of velocity, with a value of the order of
$2 - 3 \times 10^{-4}$, and with no noticeable feature at the location of the
stellar line. This measured standard deviation in time is of the expected 
order of magnitude if there is no variation of the Stokes $V$ signature, 
considering the S/N ratio in the LSD Stokes $V$ profiles listed in Table 1.
We also noted that the standard deviation of 
the null $N$ profile has the same value as that of the Stokes $V$ profile.
This stable Stokes $V$ signature implicitly requires a large scale 
magnetic field which is intrinsically stable on the timescale of our 
observations, i.e. over one year.

Finally, the Stokes $V$ signature is detected on the LSD profiles
calculated with both the full line mask and the shallow-line mask, with a very
similar shape and intensity.

A first magnetic detection was reported long ago by Babcock (1958), which 
yielded an effective longitudinal magnetic field of +270 G for neutral species, 
while ionized species yielded a null magnetic field, and the Ca~II H\&K lines
indicated -270 G. The large uncertainties in these earlier 
measurements, as re-analyzed by Preston (1969), imply that Babcock's results
must be taken as simply hinting at the presence of a magnetic field rather than
as a firm detection. This detection was not confirmed by Glagolevskij \& 
Chountonov (1998). Very recently, Hubrig et al. (2006), based on data
obtained with the FORS1 instrument on the VLT, have reported 
a weak circular polarization signal in the Ca II H\&K lines, yielding
an effective longitudinal magnetic field of the same order as the one we 
detect here on metallic lines. 

We have tried to recover Babcock's result, by performing 
LSD analysis separately on lines on neutral and ionized species. We detect a 
clear Stokes $V$ signature in both cases, with no significant difference
between them nor with the Stokes $V$ signature calculated with all lines.
We therefore do not confirm Babcock's results, neither on the amplitude of 
the field which is nearly 4 times weaker in our case, nor on the difference 
between lines of neutral and ionized species. In addition, we see no Stokes $V$ 
signal in the Ca~II H\&K lines, but the S/N ratio of our spectra is low 
in that spectral region, resulting in a noise of the order of 
$7 \times 10^{-3}$ in Stokes $V$ per spectral bin of 0.0025 nm, which would 
not allow us to detect as weak a signal as seen in the LSD average, nor the
level of signal reported by Hubrig et al. (2006).

The absence of variability of the magnetic signature in the course of one 
year can be interpreted in three different ways. Either the star is seen 
pole-on, and therefore whatever the magnetic configuration, the distribution
of longitudinal field remains always the same; or the star is seen with an
arbitrary inclination, but the magnetic field is symmetric about 
the rotation axis; or finally, the star and its magnetic field are in no 
particular configuration, but the rotation period is long enough compared to 
the one-year time span of our observations that no variation of the
Stokes $V$ or $I$ profiles can be detected; a combination of the three
interpretations is of course possible as well.
Our estimates of $v \sin i$, which can be anything between 0 and 8.6 
km\thinspace s$^{-1}$ is compatible with all three hypothesis.

A significant ambiguity about the rotation and its inclination 
angle has been discussed in the literature, in order to interpret the very low 
$v \sin i$: the star was assumed to be either rotating very slowly 
intrinsically, or to be seen pole-on. We note that there is no consensus 
about this issue: while some authors prefer the intermediate inclination
hypothesis on various and disputable theoretical grounds (Cuttela \& Ringuelet 
1990, Pogodin et al. 2005), recent interferometric observations in the 
infrared may argue in favour of a face-on configuration for the extended 
circumstellar disk of HD 190073, implying a pole-on configuration for the star
(Eisner et al. 2004). 

In order to help solving the ambiguity about the rotation of HD 190073, we 
searched for potential photometric variability of that star in 
the Hipparcos database. We find that the Hipparcos $H_p$ magnitude 
of HD 190073 shows a scatter typically of the order of 0.02 mag on a daily
time scale, with sporadic intervals where the daily scatter reaches 0.05 mag.
This scatter is significantly larger than the usual error bar associated 
with Hipparcos $H_p$ measurements, so we conclude that it must be due to 
intrinsic short term variability of the star. On the other hand, such a large
short-term scatter does not allow us to detect longer term periodicity in the
photometry that could be linked to rotation, although some hints exist for
possible periodicities around 5.5 days and 22.3 days, but with amplitudes that
are lower than the daily scatter of the data. These results are therefore
ambiguous, and no conclusion can be drawn on the rotation period of HD 190073.

\subsection{Discussion}

We now discuss the nature of the wind, circumstellar environment
and magnetic field  of HD 190073 in the light of our observational results.

First, we have seen in Sect. 3.2 that the H$\alpha$ line profile can 
be naturally explained by the
presence of a stellar wind, with a mass loss rate of the order of
$1.4 \times 10^{-8}$ M$_{\odot}$yr$^{-1}$, including a heated region
at the base of the wind. 

A basic question concerns the origin of the numerous emission lines.
Because they all correspond to lines normally expected in the 
photosphere, they must be formed in a region where the density 
is not very different from that of the photosphere, otherwise
the emission spectrum would show much more important differences with 
the photospheric absorption spectrum, in particular we would certainly
see emission lines with no photospheric absorption counterpart, and 
vice-versa there would be many more strong photospheric absorption lines 
showing no emission.

The very rich emission spectrum of HD 190073, including all lines deeper
than 0.4 in the photospheric line mask, indicates that the electron 
density of the heated region responsible for these emission lines is 
probably high, comparable to photospheric densities for this effective 
temperature and surface gravity, i.e. in the range 
$10^{13} - 10^{14}$ cm$^{-3}$. The temperature of this region must 
be significantly in excess of the effective temperature of the star, 
i.e. typically in the range 15,000 -- 20,000 K. These very rough 
estimates of the physical conditions in the heated region need to be 
verified and improved by a full non-LTE modelling of the formation of
all observed spectral lines, which is out of the scope of the present
paper.

One possible assumption is that these emission lines are formed in 
a heated region located deep at the base of the wind, where such physical
conditions could be met. We note that Cuttela \& Ringuelet 
(1990) suggest that a heated region is present at the base of the
wind, an assumption which is also supported by our tentative modelling of 
the H$\alpha$ line. 
In this heated region at the base of the wind,
which does not necessarily cover the whole stellar surface, the source
function in spectral lines is increased compared to the rest of the atmosphere.
Lines that are strong enough to be formed in this heated region at the base
of the wind have a contribution in emission, as well as a contribution in
absorption from the rest of the stellar surface. Weaker lines, which are 
formed deeper in the atmosphere, have no contribution from the heated region,
presumably located too high above the star's surface. 
A heated region at the base of the wind has
been frequently invoked to explain a few strong emission lines and lines of
superionized species in the spectra of several Herbig Ae/Be stars 
(Catala \& Talavera 1984, Catala 1988, Bouret \& Catala 1998). In the case
of HD 190073, where a much larger number of lines are seen in emission, 
the heated region just needs to be of higher density than in these
other less extreme cases.

In order to produce symmetric emission lines as observed, 
the heated region must 
have a negligible velocity gradient. This is possible if the wind acceleration
is gentle in this region, then becomes steeper higher up in the wind. 
The width of the symmetric, more or less gaussian-shaped emission lines
may be attributed to turbulent motions
inside the heated region. A turbulent velocity of the order of 39
km\thinspace s$^{-1}$ would account for the observed FWHM of
65 km\thinspace s$^{-1}$ of these lines. This turbulent 
velocity would be supersonic in the physical conditions of this
heated region ($T \approx 15,000 - 20,000$ K). We note that supersonic
turbulence has often been reported in the winds of hot stars, in 
particular in the case of PMS Herbig Ae/Be stars (see e.g. Bouret \& Catala
1998).

Although it is attractive, one problem with the interpretation of the emission 
lines in terms of a heated region at the base of the wind is that we would 
expect the emission lines to be centered in the star's rest frame, or even
blueshifted if the outflow velocity in this region is not 
negligible, whereas we observe instead a variable but systematic redshift 
of the emission lines with respect to 
the photospheric absorption lines. Therefore, if
this interpretation is correct, a more complex velocity pattern than a
simple spherically symmetric wind must be assumed to account for this 
behaviour.

For instance, the heated region could be related to an accretion flow 
channelling matter from a circumstellar disk onto the star's surface in one
or several  magnetic funnels, in a similar way as what has been
proposed for classical T Tauri stars (Paatz \& Camenzind 1996, 
Calvet \& Gullbring 1998, Romanova
et al. 2003). The shocks produced by these accretion columns may be sufficient 
to heat a sufficiently large region at the star's surface to produce the 
observed emission lines, which require temperatures and particle
densities in the range 15,000 -- 20,000 K and $10^{13} - 10^{14}$ cm$^{-3}$,
respectively. Extrapolating the calculations of the accretion 
shock structure in classical T Tauri stars (Calvet \& Gullbring 1998) 
to the case of HD 190073, we suggest that these conditions can probably
be met either in the preshock region, or in the photospheric region underlying
the shocks. In the first hypothesis, a strong accretion rate is needed
to provide the high particle densities required for the formation of the
observed emission lines: still extrapolating the results of Calvet \& Gullbring
(1998), and assuming a filling factor of the order of 1\% for the shocked 
regions at the star's surface, we find that accretion rates of several 
$10^{-7}$ M$_{\odot}$yr$^{-1}$ are needed. In the second hypothesis, weaker
accretion rates would probably be sufficient, although a full modelling of
the photosphere perturbed by the shock, which is outside the scope of 
this paper, would be necessary to study this issue. 
 
Although the above interpretation of the line emission components is 
attractive, their width and their small variable redshift with respect to the 
absorption components remain to be explained. We may speculate that the 
kinetic energy of the accretion flow in the preshock region is partially 
transformed into turbulent energy, through flow instabilities, in order to 
explain the width of the emission components. On the other hand, the material
being almost free falling in the preshock region, we would expect a much larger 
redshift than observed for the emission components, unless the angle between
the accretion flow and the line of sight is systematically close to 
$90^{\circ}$. For instance, a simple configuration with an inclination angle 
$i \approx 0^{\circ}$ for the star's rotation axis (star seen pole-on) 
and an angle 
$\beta \approx 90^{\circ}$ between the rotation axis and the axis of a dipolar 
magnetic field, with matter being accreted onto the magnetic poles, would
be compatible with the low $v \sin i$ and the width and redshift of
the emission components if they are formed in the preshock region. 

A detailed modelling of accretion flows in intermediate mass PMS stars, 
analogous to the work of Calvet \& Gullbring (1998) for classical T Tauri
stars, is clearly needed in the future to study these issues in more details.



Alternative scenarios can be invoked for the formation of these emission
lines. One possibility is that the emission lines are formed at the 
surface of a circumstellar disk, and are broadened by the disk's Keplerian 
rotation. In this case, we expect the emission components to exhibit
a characteristic double-peak shape, which may resemble the observed profiles,
for some particular choice of disk parameters. 
A Keplerian velocity at the surface of $V_K^* = 390$ km\thinspace s$^{-1}$
can be derived using the mass and radius of the star previously estimated,
and the observed width of the emission lines (65 km\thinspace s$^{-1}$ FWHM)
implies that the inclination angle $i$ between the disk axis and the line of 
sight is such that $sin$i$ = 0.083 \times (r_{em}/R_*)^{1/2}$, where $R_*$ and 
$r_{em}$ are respectively the stellar radius and the radius of the region
of formation of the emission lines. This scenario would therefore imply
that the system of HD 190073 and its disk is seen at small inclination, for
reasonable values of $r_{em}$ of a few stellar radii.
The emission lines can be formed for instance by scattering of the stellar 
flux in an optically thin inner region of the disk, in a geometry similar to
that suggested by Vink et al. (2005) for several other pre-main sequence 
stars.
An important difficulty with this assumption, though, is that the emission 
lines all correspond to lines present in the photosphere, so that their 
region of formation in the disk must have density and temperature conditions
which are not very different from those of the photosphere, as argued earlier.
Such a coincidental similarity between photospheric and disk conditions is 
certainly unlikely. Also the systematic redshift between emission lines and
photospheric absorptions remains unexplained in this scenario. 

The data presented in this paper are clearly not sufficient to conclude
on the formation of the observed emission components. 
More detailed studies would be needed to 
investigate the various hypothesis presented above. 
In particular further high quality 
spectropolarimetric data, and a full NLTE radiative transfer modelling
of the observed lines should help us understand how these lines are formed, 
and constrain the configuration of the star, its magnetic field and 
circumstellar environment.

\section{Conclusion}

Our discovery of a globally structured magnetic field in HD 190073, a 
young 2.85 $M_{\odot}$ Herbig Ae star, brings
a piece to the puzzle of stellar magnetism for intermediate mass stars. 

Our estimate of 1.2 Myrs for its age indicates that stars less massive 
than 3 $M_{\odot}$ can display significant surface magnetic fields at a 
very young age, in contradiction 
with Hubrig et al's (2000) conclusion that such stars become magnetic only 
after they have completed 30\% of their main sequence life.

The star does not show the usual strong chemical peculiarities of Ap stars, 
possibly indicating that they develop on a time scale longer than 1 Myr, 
or that the accretion/mass loss processes that HD 190073 is currently 
experiencing are sufficient to disrupt their formation.

We were not able to constrain the geometry of HD 190073 and its magnetic field,
due to the absence of variability of the polarimetric signature.
Clearly a longer term investigation of this star should be undertaken.

Further spectropolarimetric monitoring of additional Herbig Ae and Be stars 
should also be
performed, in order to detect other magnetic stars among them and to constrain
their magnetic geometries and intensities. Of great importance would be a
statistical study of magnetism and rotation in pre-main sequence stars, in 
order to understand the evolution of magnetic fields and angular momentum 
during this early phase.

\acknowledgements{We warmly thank the CFHT staff for their efficient help 
during the observations. We are grateful to the referee, Dr. C.M. Johns-Krull,
for his very useful comments which helped us improve this paper. 
GAW acknowledges support from the Academic Research 
Programme of the Department of National Defence (Canada). GAW and JDL 
acknowledge support by the Natural Sciences and Engineering Council of Canada.}

{}


\begin{thebibliography}{}

\bibitem{}
Acke, B., van den Ancker, M.E., Dullemond, C.P. 2005, A\&A 436, 209

\bibitem{}
Acke, B., Waelkens, C. 2004, A\&A 427, 1009

\bibitem{}
Babcock, H.W. 1958, ApJS 3, 141

\bibitem{}
Bouret, J.-C., Catala, C. 1998, A\&A 340, 163

\bibitem{}
Calvet, N., Gullbring, E. 1998, ApJ 509, 802

\bibitem{}
Catala, C., Talavera, A. 1984, A\&A 140,421 

\bibitem{}
Catala, C.  1988, A\&A 193, 222 

\bibitem{}
Catala, C., et al. 1999, A\&A 345, 884 (1999) 

\bibitem{}
Chen, P. S., Wang, X. H., He, J. H. 2000, Ap\&SS, 271, 259

\bibitem{}
Cidale, L., Zorec, J., Morrell, N. 2000, in IAU Colloq. 175, The Be
Phenomenon in Early-Type Stars, ASP Conf. Ser., 214, 87

\bibitem{}
Cuttela, M., Ringuelet, A.E. 1990, MNRAS 246, 20

\bibitem{}
de Winter, D., van den Ancker, M. E., Maira, A., et al. 2001, A\&A, 380, 609


\bibitem{}
Donati, J.-F., Semel, M., Carter, B.D., Rees, D.E., Cameron, A.C. 1997, MNRAS
291, 658

\bibitem{}
Eisner, J.A., Lane, B.F., Hillenbrand, L.A., Akeson, R.L., Sargent, A.I. 2004, 
ApJ 613, 1049

\bibitem{}
Glagolevskij, Y. V.,  Chountonov, G. A. 1998, Bull. Special
Astrophys. Obs., 45, 105

\bibitem{}
Grady, C. 2005, in The Nature and Evolution of Disks Around Hot Stars, 
R. Ignace and K. G. Gayley (eds), ASP Conference Series, Vol. 337, p.155

\bibitem{}
Gray, D.F. 1992, The observation and analysis of stellar photospheres 
(Cambridge Astrophysics Series, Cambridge: Cambridge University Press, 
1992, 2nd ed., ISBN 0521403200.)

\bibitem{}
Hubrig, S., Sch\"oller, M., Yudin, R.V. 2004, A\&A 428, L1

\bibitem{}
Hubrig, S., Yudin, R.V., Sch\"oller, M., Pogodin, M.A. 2006, A\&A 446, 1089


\bibitem{}
K\"onigl, A. 1991, ApJ 370, L39

\bibitem{}
Landstreet, J.D. 1998, A\&A 338, 1041

\bibitem{}
Malfait, K., Bogaert, E., Waelkens, C. 1998, A\&A 331, 211

\bibitem{}
Mestel, L. 2001, ASP Conf. Series, vol. 205, p. 3

\bibitem{}
Mihalas D., Kunasz P. 1978, ApJ 219, 635

\bibitem{}
Moss, D. 2001, ASP Conf. Series, vol. 248, p. 305

\bibitem{}
Morel, P. 1997, A\&AS 124, 597

\bibitem{}
Paatz, G., Camenzind, M. 1996, A\&A 308, 77

\bibitem{}
Palla, F., Stahler, S. 1990, ApJ 360, L47

\bibitem{}
Pogodin, M., Miroshnichenko, A.S., Tarasov, A.E., et al. 2004, A\&A 417, 715

\bibitem{}
Pogodin, M., Franco, G.A.P., Lopes, D.F. 2005, A\&A 438, 239

\bibitem{}
Praderie, F., Simon, T., Catala, C., Boesgaard, A.M. 1986, ApJ 303, 311 

\bibitem{}
Reiners A., Schmitt J.H.M.M., 2003, A\&A 398, 647

\bibitem{}
Romanova, M. M., Ustyugova, G. V., Koldoba, A. V., Wick, J. V., 
Lovelace, R. V. E. 2003, ApJ 595, 1009


\bibitem{}
Van den Ancker, M.E., De Winter, D., Tjin A Djie, H.R.E. 1998, A\&A 330, 145

\bibitem{}
Vink, J., Drew, J.E., Harries, T.J., Oudmaijer, R.D., Unruh, Y. 2005, MNRAS
359, 1049

\bibitem{}
Wade, G.A., Drouin, D., Bagnulo, S., Landstreet, J.D., Mason, E., Silvester, 
J., Alecian, E., B\"ohm, T., Bouret, J.-C., Catala, C., Donati, J.-F. 2005,
A\&A, 442, L31

\end{thebibliography}
\end{document}